\documentclass{article}
\usepackage{spconf,amsmath,graphicx,hyperref}

\usepackage{multirow}
\usepackage{arydshln}

\usepackage[markup=default]{changes}

\usepackage{natbib}
\setcitestyle{numbers,square,comma}

\title{SFQA: A Comprehensive Perceptual Quality Assessment Dataset for Singing Face Generation}
\name{Zhilin Gao$^{1,*}$, Yunhao Li$^{1,*}$, Sijing Wu$^{1,\dagger}$, Yucheng Zhu$^{1}$, Huiyu Duan$^{1}$, Guangtao Zhai$^{1,\dagger}$\thanks{$^{*}$ Equal contribution. $^{\dagger}$ Corresponding author. \newline \indent This work was supported by the National Natural Science Foundation of China under Grant 62571324}}
\address{$^{1}$Shanghai Jiao Tong University, Shanghai, China} 

\begin{document}
%
\maketitle
\begin{abstract}  
The Talking Face Generation task has enormous potential for various applications in digital humans and agents, etc. Singing, as a common facial movement second only to talking, can be regarded as a universal language across ethnicities and cultures. However, it is often underestimated in the field due to lack of singing face datasets and the domain gap between singing and talking in rhythm and amplitude. More significantly, the quality of Singing Face Generation (SFG) often falls short and is uneven or limited by different applicable scenarios, which prompts us to propose timely and effective quality assessment methods to ensure user experience. To address existing gaps in this domain, this paper introduces a new SFG content quality assessment dataset SFQA, built using 12 representative generation methods. During the construction of the dataset, 100 photographs or portraits, as well as 36 music clips from 7 different styles, are utilized to generate 5,184 singing face videos that constitute the SFQA dataset. To further explore the quality of SFG methods, subjective quality assessment is conducted by evaluators, whose ratings reveal a significant variation in quality among different generation methods. Based on our proposed SFQA dataset, we comprehensively benchmark the current objective quality assessment algorithms.
\end{abstract}

\section{Introduction}
\label{sec:intro}

In recent years, Artificial Intelligence Generated Content (AIGC) has opened up new possibilities for various aspects of life and artistic creation, finding widespread application across industries such as business, film, television, and entertainment. Among them, \textbf{Talking Face Generation} (TFG, \emph{i.e.} Audio-driven Portrait Image Animation) has attracted increasing attention due to its broad application prospects in digital embodied intelligence, and related technologies are also developing rapidly \cite{li2025samr, wu2023ganhead}. Nowadays, a lively and realistic talking face video can usually be generated with just one image and one audio clip. Despite these advances, the quality of most TFG content currently varies, as illustrated in Fig.~\ref{fig2}. These TFG methods have varying degrees of applicability to different scenarios and often struggle to meet users' expectations. Singing, as another common facial movement, plays a significant role in art and entertainment. Despite fundamental differences between singing and speaking in rhythm, expression, and amplitude, both domains suffer from analogous quality degradation issues. This compels the creation of singing-specific datasets and development of corresponding quality assessment methodologies \cite{wu2024mmhead, wu2023singinghead}.

\begin{figure}[t]
\centering
\includegraphics[width=\columnwidth]{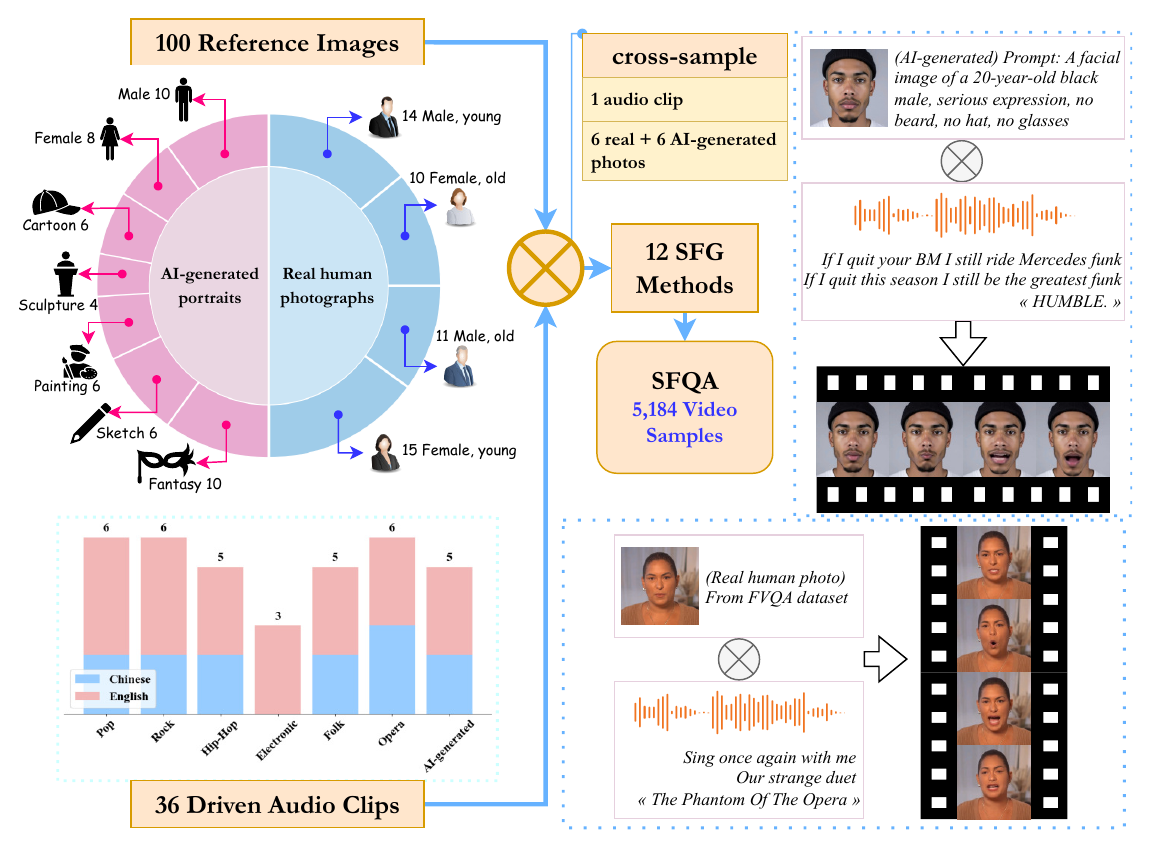}
\vskip -0.03in
\caption{The construction process of SFQA dataset and diverse raw materials.} 
\label{fig1}
\vskip -0.2in
\end{figure}

To address the challenge of assessing \textbf{Singing Face Generation} (SFG) content, we develop a comprehensive quality assessment dataset, SFQA. As shown in Tab.~\ref{tab1}, this dataset consists of 12 representative generative algorithms. By cross-sampling 100 portraits and 36 music clips based on the category division in Fig.~\ref{fig1}, 5,184 singing face videos are successfully generated. Volunteers are then invited to assess the quality of the SFG content within the dataset. The results of subjective evaluations reveal significant quality variations and differences in style tendencies and quality defects across the different generative algorithms. These findings emphasize the critical importance of quality assessment for SFG content. Furthermore, this study tests several commonly used objective quality assessment algorithms to benchmark their performance on SFQA dataset. The experimental results demonstrate the limitations of existing objective assessment methods and highlight the need for the development of more targeted assessment algorithms.

\begin{table}[htbp]
\small
\centering
\renewcommand\tabcolsep{5pt}
\vskip -0.1in
\caption{generation method list}
\resizebox{0.95\linewidth}{!}
{
\begin{tabular}{c|c|c|c}
\hline
Label & Method & Year & Backbone Architecture \\
\hline
SadT & SadTalker \cite{zhang2022sadtalker} & 2023 & Pre-trained Network + VAE \\
AniP & AniPortrait \cite{wei2024aniportrait} & 2024 & Projection + Diffusion \\
VExp & V-Express \cite{wang2024V-Express} & 2024 & Stable Diffusion v1.5 \\
hallo2 & hallo2 \cite{cui2024hallo2} & 2024 & Latent Diffusion Model \\
FLOAT & FLOAT \cite{ki2024float} & 2024 & flow matching \\
memo & memo \cite{zheng2024memo} & 2024 & Latent Diffusion Model \\
EchoM & EchoMimic \cite{chen2024echomimic} & 2025 & Stable Diffusion + VAE \\
sonic & sonic \cite{ji2025sonic} & 2025 & Diffusion Model \\
hallo3 & hallo3 \cite{cui2024hallo3} & 2025 & Diffusion Transformer \\
SkyR & SkyReels-A1 \cite{qiu2025skyreels} & 2025 & Diffusion Transformer \\
fantasy & fantasy-talking \cite{wang2025fantasytalking} & 2025 & Latent Diffusion Model \\
DICE & DICE-Talk \cite{tan2025dicetalk} & 2025 & Stable Video Diffusion \\

\hline
\end{tabular}
}
\vskip -0.17in
\label{tab1}
\end{table}

\begin{figure}[hbtp]
\begin{minipage}[b]{0.49\linewidth}
  \centering
  \centerline{\includegraphics[width=1.0\textwidth]{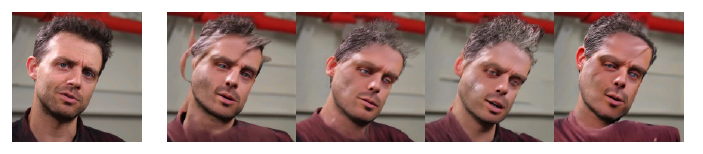}}
  \vspace{-2mm}
  \centerline{(a) Distortion - face}\medskip
\end{minipage}
\hfill
\begin{minipage}[b]{0.49\linewidth}
  \centering
  \centerline{\includegraphics[width=1.0\textwidth]{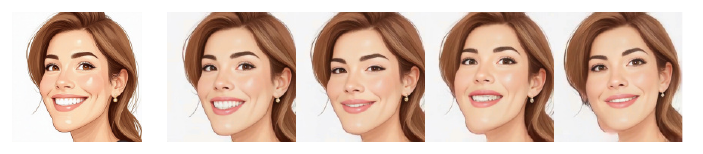}}
  \vspace{-2mm}
  \centerline{(b) Distortion - jaw}\medskip
\end{minipage}

\begin{minipage}[b]{0.49\linewidth}
  \centering
  \centerline{\includegraphics[width=1.0\textwidth]{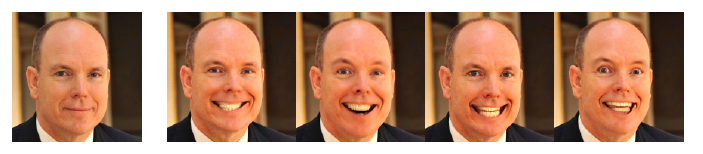}}
  \vspace{-2mm}
  \centerline{(c) Distortion - teeth}\medskip
\end{minipage}
\hfill
\begin{minipage}[b]{0.49\linewidth}
  \centering
  \centerline{\includegraphics[width=1.0\textwidth]{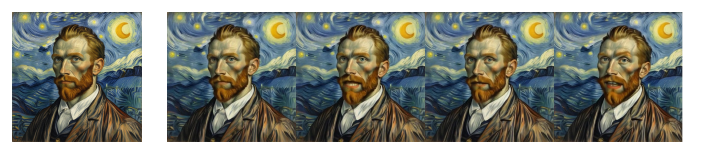}}
  \vspace{-2mm}
  \centerline{(d) Noise}\medskip
\end{minipage}

\begin{minipage}[b]{0.49\linewidth}
  \centering
  \centerline{\includegraphics[width=1.0\textwidth]{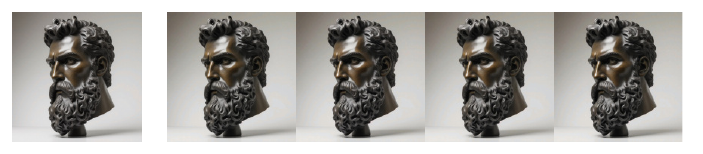}}
  \vspace{-2mm}
  \centerline{(e) Still, lip mismatch}\medskip
\end{minipage}
\hfill
\begin{minipage}[b]{0.49\linewidth}
  \centering
  \centerline{\includegraphics[width=1.0\textwidth]{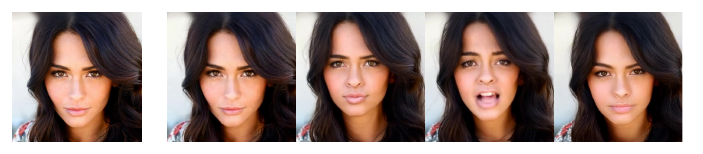}}
  \vspace{-2mm}
  \centerline{(f) Sudden changes}\medskip
\end{minipage}

\begin{minipage}[b]{0.49\linewidth}
  \centering
  \centerline{\includegraphics[width=1.0\textwidth]{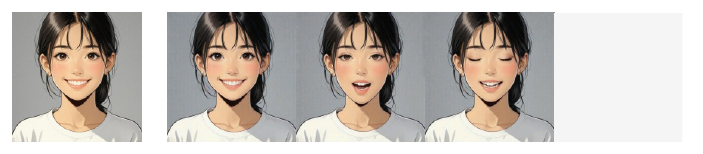}}
  \vspace{-2mm}
  \centerline{(g) Truncation}\medskip
\end{minipage}
\hfill
\begin{minipage}[b]{0.49\linewidth}
  \centering
  \centerline{\includegraphics[width=1.0\textwidth]{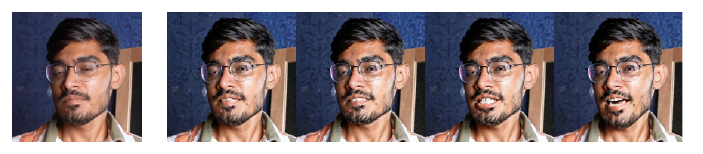}}
  \vspace{-2mm}
  \centerline{(h) Abnormal color rendering}\medskip
\end{minipage}

\begin{minipage}[b]{0.49\linewidth}
  \centering
  \centerline{\includegraphics[width=1.0\textwidth]{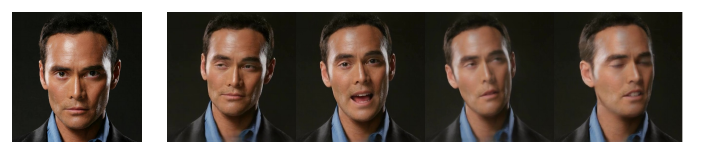}}
  \vspace{-2mm}
  \centerline{(i) Blur}\medskip
\end{minipage}
\hfill
\begin{minipage}[b]{0.49\linewidth}
  \centering
  \centerline{\includegraphics[width=1.0\textwidth]{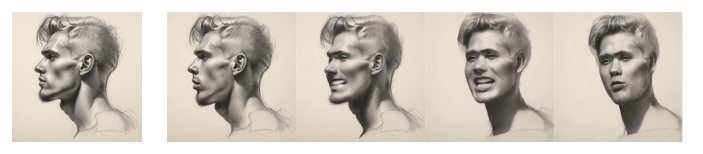}}
  \vspace{-2mm}
  \centerline{(j) Changes in character features}\medskip
\end{minipage}

\vskip -0.15in
\caption{Typical quality issues in contemporary TFG (Left: original reference image example)}
\label{fig2}
\vskip -0.15in
\end{figure}

\section{Related Work}
\label{sec:related}

\subsection{Talking Face Generation}
\label{ssec:related1}

The task of Talking Face Generation (TFG) has attracted a lot of attention due to its potential for the development of digital human generation \cite{ji2025sonic} and embodied agents. Many researchers are dedicated to developing methods to generate high-quality synchronized facial expressions driven by speech. Techniques such as flow matching \cite{ki2024float} and diffusion models \cite{qiu2025skyreels, zheng2024memo} have demonstrated state-of-the-art performance. Some methods also attempt to extend facial movements to the half body or the whole body, incorporating gesture information beyond facial expressions \cite{Meng_2025_EchoMimicV2}. However, the wide variety of driving vocal audio has led to many refined subsets in this field, such as singing face generation, which has gradually attracted the attention of researchers due to the diverse impact of music on speech. This article focuses mainly on this subset, aiming to offer valuable insights for the continuous development and improvement of these technologies.

\subsection{Quality Assessment Method}
\label{ssec:related2}

The current quality assessment methods for AIGC can be divided according to modality, including single modality quality assessment (image \cite{li2025aghi}, audio, video \cite{sun2022simplevqa, wu2022fastvqa}, etc.) and cross-modality quality assessment (audio-visual \cite{gao2025ges}, text-visual, etc.). From a methodological perspective, it can be roughly divided into deep learning based methods, Large Language Model (LLM) based methods \cite{zhou2026mi3s}, and synchronization methods \cite{Chung16syncnet}. For TFG tasks, some consider it as a local area evaluation of dynamic digital humans \cite{wusijing2025fvq}, while others view it as a 2D rendered animated video and directly use audio-visual quality assessment methods \cite{zhou2025better}. Compared to conventional AIGC, the significant characteristics of TFG content remain substantially underexplored in recent quality assessment research, including subject consistency of human, speech-action desynchronization, and diverse classification of raw materials, which are all issues that our proposed SFQA dataset aims to address in this paper.

\section{Dataset Construction}
\label{sec:dataset}

\subsection{Material Collection} 
\label{ssec:data1}

The current SFG methods typically necessitate a reference image and a driving audio clip as the raw materials for generation. To ensure content diversity within the constructed dataset, we selected 50 photographs of real human faces and 50 AI-generated human or humanoid portraits as reference images and 36 music clips as driving audio. As illustrated in Fig.~\ref{fig1}, real person photos are sourced from two distinct datasets: FVQA \cite{wusijing2025fvq} and CelebAMask-HQ \cite{CelebAMask-HQ}, and these 50 images are categorized into four groups according to gender and age. AI-generated pictures are produced using Stable Diffusion 3.5 Large Turbo \cite{sd3.5} and can be divided into seven categories according to the content of the prompt words. The 36 music clips contain Chinese and English languages and can be categorized into seven distinct styles, with several clips originating from AI-generated compositions utilizing LeVo \cite{lei2025levo}. All images have undergone pre-processing to achieve a uniform resolution of 1024 $\times$ 1024 pixels, while the duration of the audio clips ranges from 5 to 7 seconds, ensuring the inclusion of complete sentences. These images and audio are systematically sampled to ensure that the input image-audio pairs encompass all category combinations, maintaining proportions in terms of quantity. The above content suggests that the SFQA dataset possesses a rich and diverse array of singing face content, reflecting its extensive coverage across various categories.

\begin{figure}[t]
\centering
\includegraphics[width=0.5\textwidth]{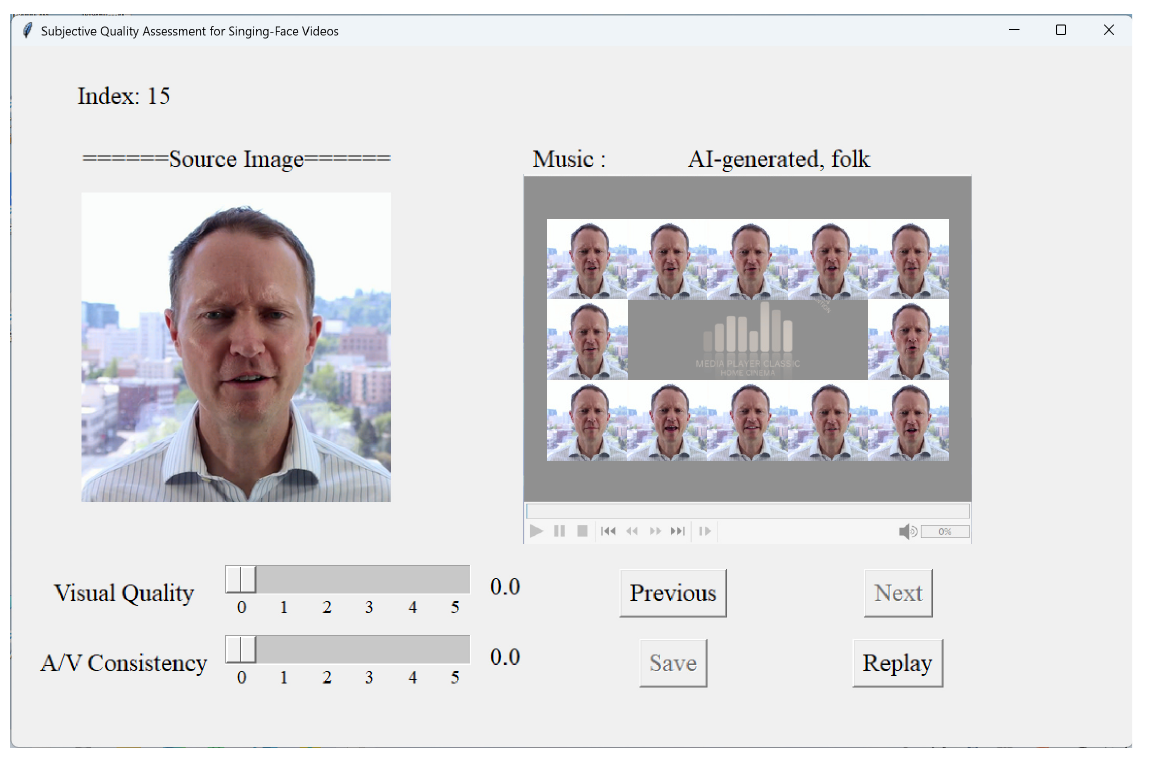}
\vskip -0.2in
\caption{User-rating GUI for subjective evaluation.}
\label{fig3}
\vskip -0.2in
\end{figure}

\subsection{Selected Generation Methods} 
\label{ssec:data2}

As shown in Tab.~\ref{tab1}, 12 representative SFG methods are selected to generate video samples from raw materials using their default weights and code. The selection of these 12 methods mainly considers their influence in the field and the coverage of different architectures. It is important to note that methods such as AniP\cite{wei2024aniportrait}, SkyR\cite{qiu2025skyreels} and EchoM\cite{chen2024echomimic} impose limitations on the resolution of input images, which consequently leads to a degradation in visual quality. Preliminary observation of generated video samples reveal distinct tendencies and defects exhibited by the aforementioned SFG methods. For instance: DICE \cite{tan2025dicetalk} produces dental artifacts during tooth exposure. Fantasy \cite{wang2025fantasytalking} exhibits premature frame truncation of terminal frames, which affects audio continuity. And multiple approaches display audio-lip de-synchronization and artifacts including background warping, noise artifacts, and chromatic aberrations to varying degrees. Certain issues are illustrated in Fig.~\ref{fig2}. Overall, existing SFG methods exhibit significant limitations in terms of quality.

\subsection{Subjective Experiment}
\label{ssec:data3}

To further assess the quality of SFG content, subjective scoring experiments are conducted using the established SFQA dataset. 18 participants are invited to subjectively rate the singing face videos. The experimental design employs absolute category ratings to assess two key aspects: visual perceptual quality and audio-visual consistency. The former evaluates the visual effects of the video, including noise, blur, distortion, motion coherence, and consistency of identity features (hairstyle, skin color, facial features). The latter evaluates the consistency between music clips and lip movements, as well as audio truncation caused by the SFG methods.

These data are collected via our custom interface, as illustrated in Fig.\ref{fig3}, which displays a reference image, audio clip information, the corresponding video sample, and two 5-point rating sliders to collect scores. Prior to formal deployment, subjects must receive training to become acquainted with the rating criteria. At the end of the experiment, a total of $31,104=5,184 \times 6$ subjective ratings are collected. These scores will be scaled into z-scores, and after excluding unreliable ratings, the final mean opinion scores (MOSs) for each sample will be determined by calculating the average of the remaining z-scores.

\begin{figure}[hb]

\begin{minipage}[b]{0.49\linewidth}
  \centering
  \centerline{\includegraphics[width=1.0\textwidth]{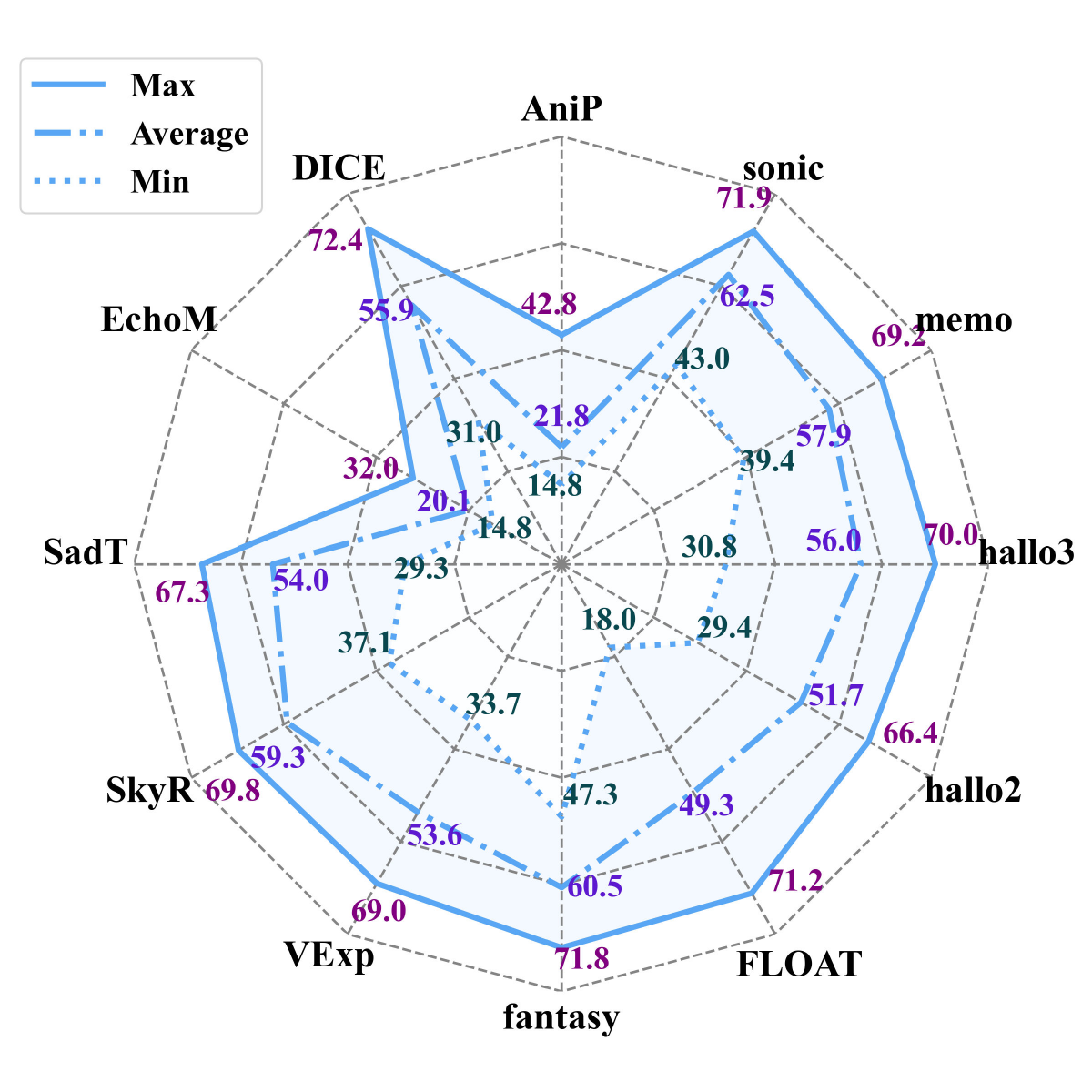}}
\end{minipage}
\hfill
\begin{minipage}[b]{0.49\linewidth}
  \centering
  \centerline{\includegraphics[width=1.0\textwidth]{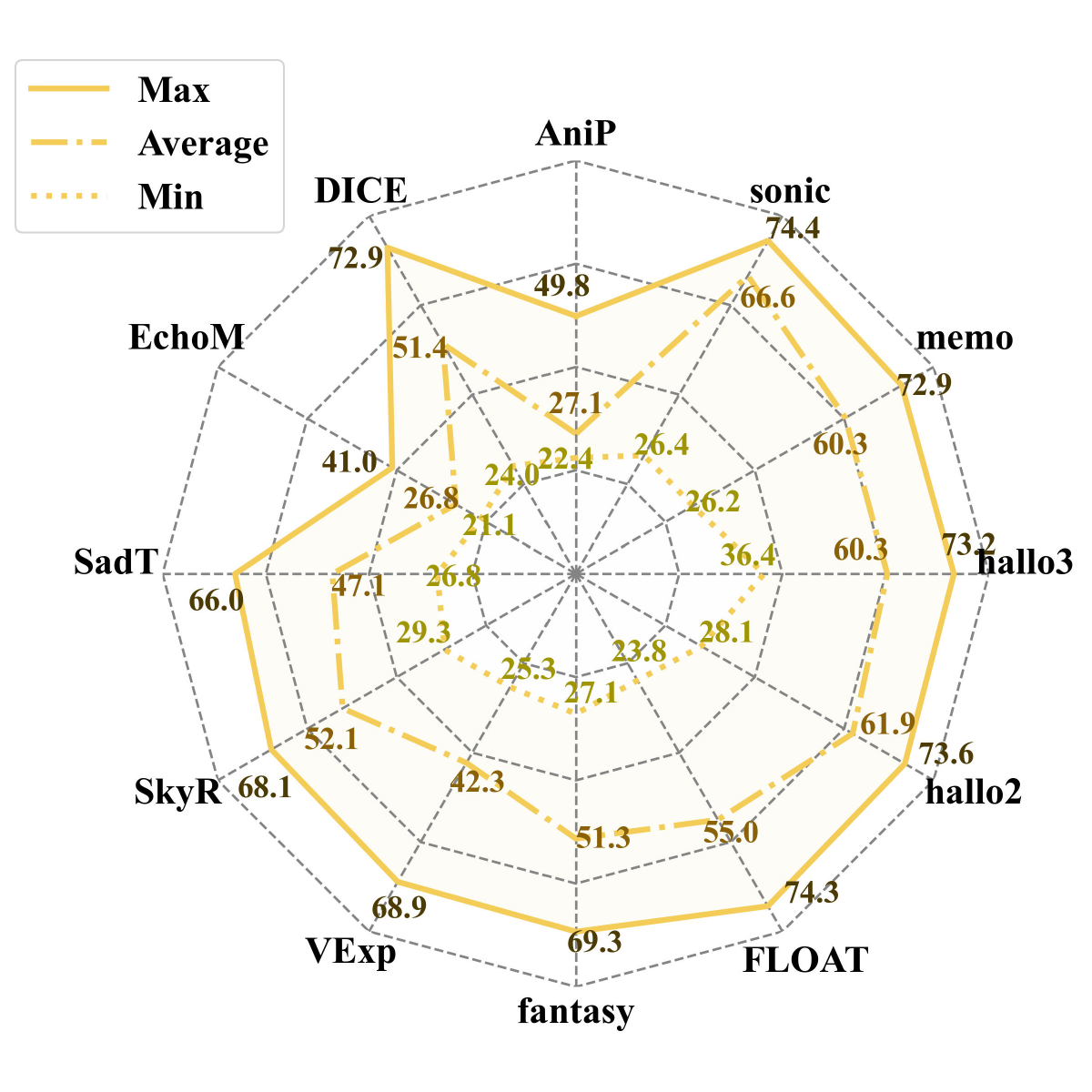}}
\end{minipage}
\centerline{(a) MOS comparison across SFG approaches.}\medskip

\begin{minipage}[b]{0.49\linewidth}
  \centering
  \centerline{\includegraphics[width=1.0\textwidth]{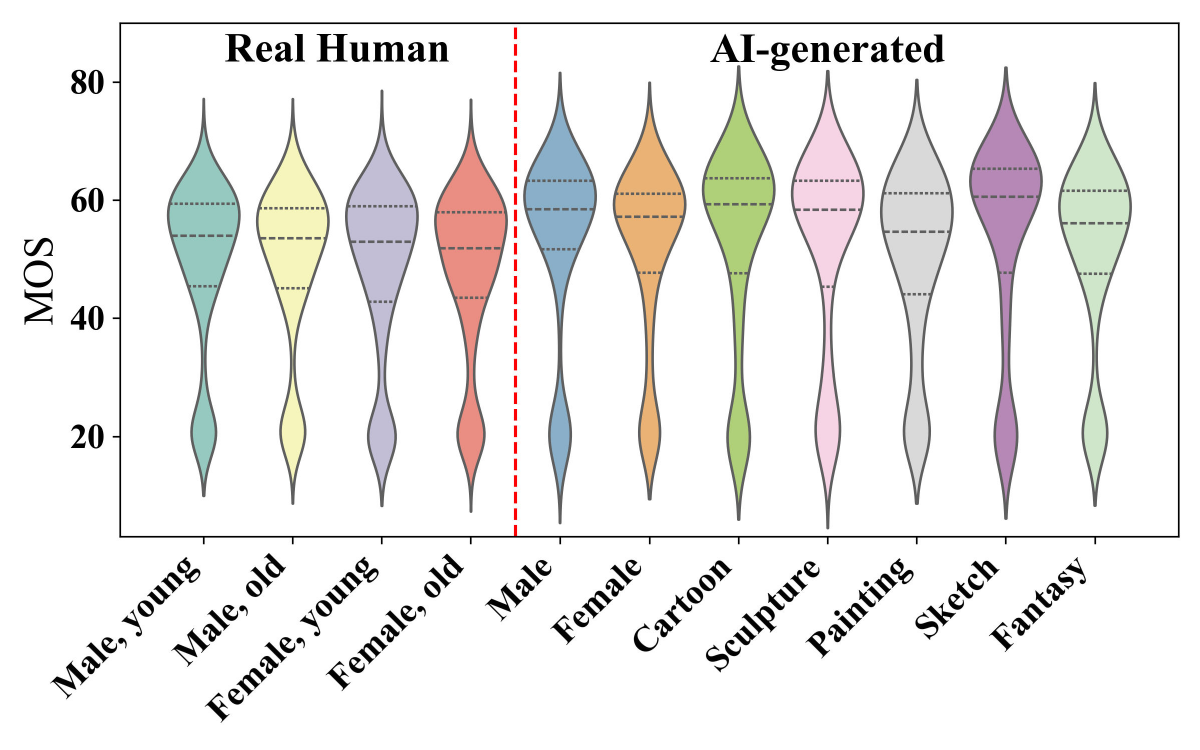}}
\end{minipage}
\hfill
\begin{minipage}[b]{0.49\linewidth}
  \centering
  \centerline{\includegraphics[width=1.0\textwidth]{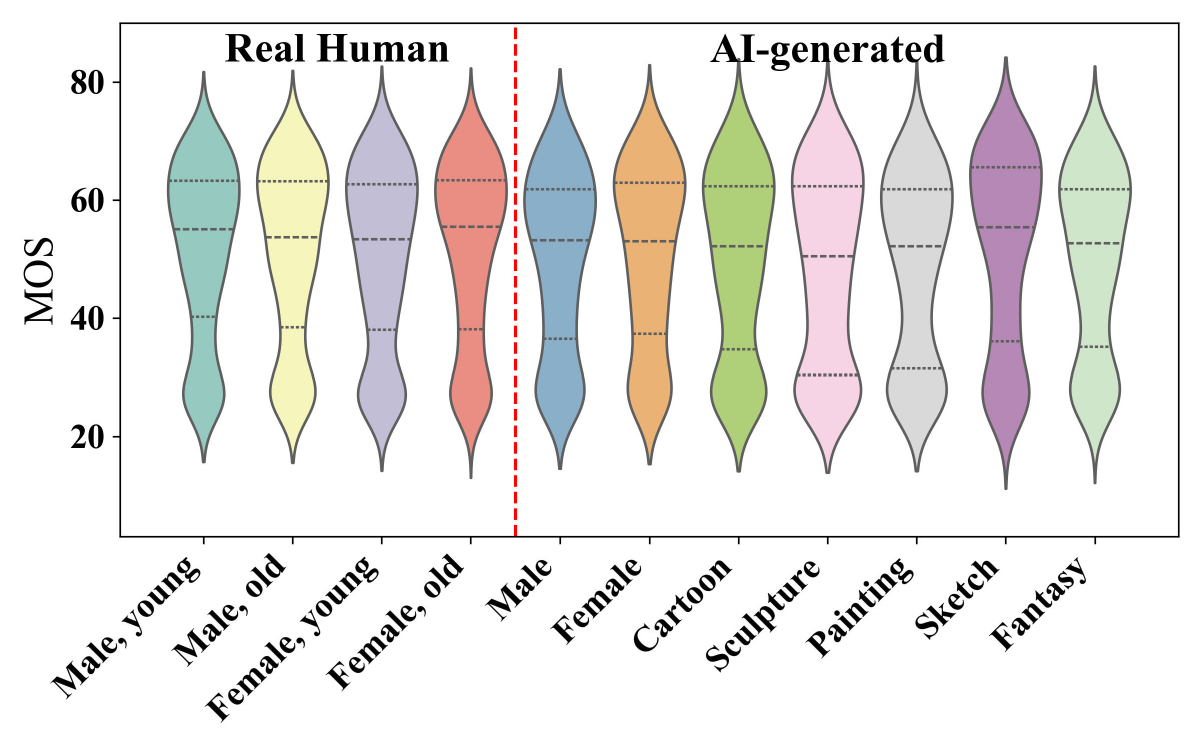}}
\end{minipage}
\centerline{(b) MOS comparison of reference image categories.}\medskip

\begin{minipage}[b]{0.49\linewidth}
  \centering
  \centerline{\includegraphics[width=1.0\textwidth]{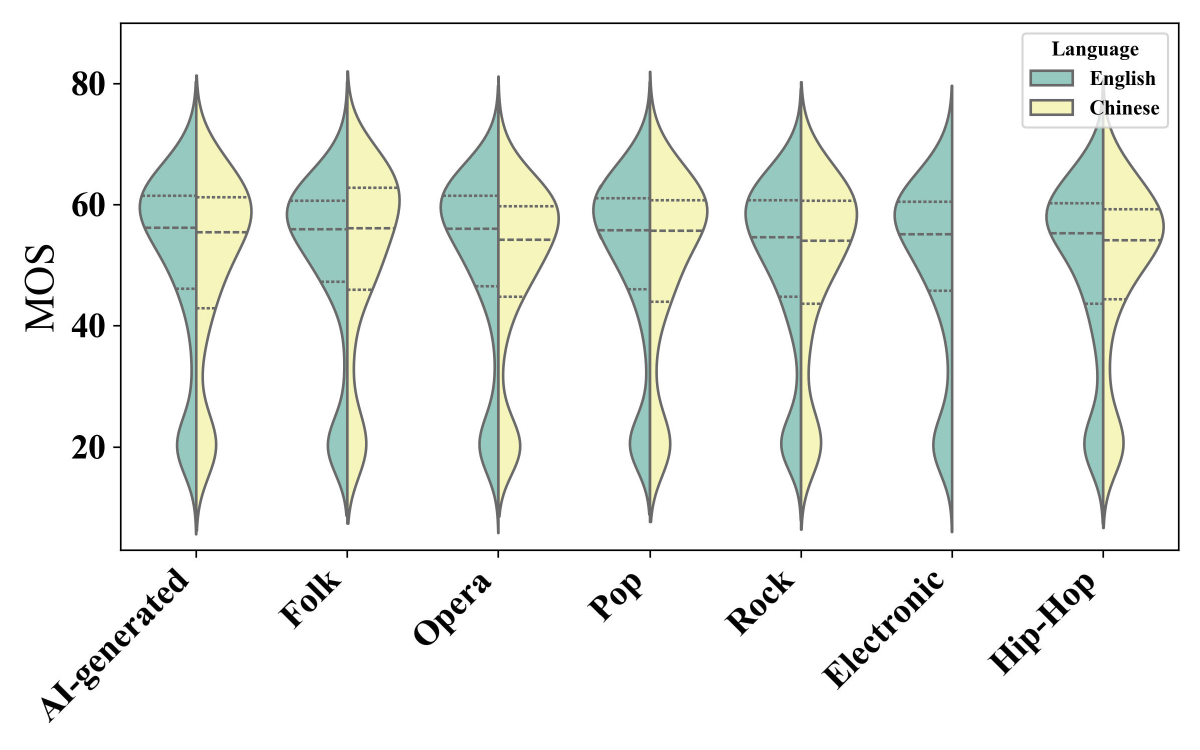}}
\end{minipage}
\hfill
\begin{minipage}[b]{0.49\linewidth}
  \centering
  \centerline{\includegraphics[width=1.0\textwidth]{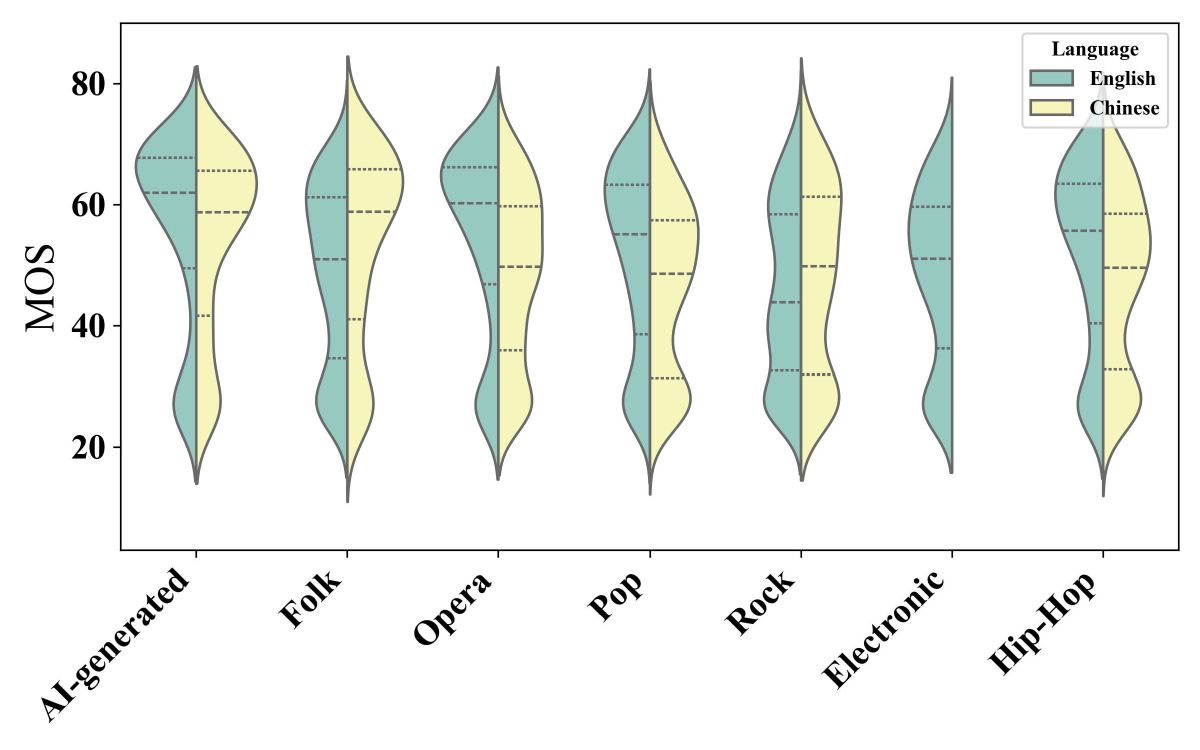}}
\end{minipage}
\centerline{(c) MOS comparison of driven audio styles.}\medskip

\caption{The distribution of subjective scores: (Left) Visual Perception Quality • (Right) Audio-Visual Consistency}
\label{fig4}
\vskip -0.27in
\end{figure}

\begin{table*}[tbp]
\small
\centering
\renewcommand\arraystretch{1.03}
\renewcommand\tabcolsep{4pt}
\vskip -0.1in
\caption{Performance comparisons of several standard quality assessment methods on the SFQA dataset. Best results are bolded, second-best are underlined.}
\resizebox{\linewidth}{!}
{
\begin{tabular}{cl|ccc|ccc|ccc}
\hline
\multicolumn{2}{c|}{Dimension} & \multicolumn{3}{c|}{Visual Perceptual Quality} & \multicolumn{3}{c|}{Audio-Visual Consistency} & \multicolumn{3}{c}{Overall Average} \\
\hdashline
Model Type & Model & SRCC & PLCC & KRCC & SRCC & PLCC & KRCC & SRCC & PLCC & KRCC \\
\hline

\multirow{2}{*}{\shortstack{Audio-Visual \\ LMMs}}& VideoLLaMA2.1-7B-AV \cite{damonlpsg2024videollama2} & 0.4322 & 0.6322 & 0.3528 & 0.2378 & 0.4080 & 0.1710 & 0.3350 & 0.5201 & 0.2484\\
~ & Qwen2.5-Omni-3B \cite{xu2025qwen25omnitechnicalreport} & 0.1634 & 0.3056 & 0.1308 & 0.0999 & 0.3216 & 0.0758 & 0.1317 & 0.3136 & 0.1033  \\
\hdashline

\multirow{2}{*}{Syncnet} & LSE-C \cite{Chung16syncnet} & 0.0965 & 0.2549 & 0.0758 & 0.4952 & 0.5317 & 0.3420 & 0.2959 & 0.3933 & 0.2089 \\
~ & LSE-D \cite{Chung16syncnet} & 0.1634 & 0.3056 & 0.1308 & 0.1000 & 0.3216 & 0.0758 & 0.1317 & 0.3136 & 0.1033 \\
\hdashline

~ & SimpleVQA \cite{sun2022simplevqa} & 0.6462 & 0.7736 & 0.4721 & 0.7272 & 0.7799 & 0.5356 & 0.6867 & 0.7768 & 0.5039 \\
VQA & FAST-VQA \cite{wu2022fastvqa} & 0.7165 & 0.7967 & 0.5323 & \underline{0.8189} & \underline{0.8605} & \underline{0.6241} & \underline{0.7677} & \underline{0.8286} & \underline{0.5782} \\
~ & DOVER \cite{wu2023dover} & \underline{0.7258} & \underline{0.8125} & \underline{0.5444} & 0.8012 & 0.8298 & 0.6072 & 0.7635 & 0.8212 & 0.5758 \\
\hdashline

\multirow{4}{*}{\shortstack{Multi-modal \\ Alignment}}& AVID-CMA \cite{avid_cma} & 0.1258 & 0.1766 & 0.0850 & 0.3703 & 0.3819 & 0.2524 & 0.2481 & 0.2793 & 0.1687 \\
~ & VAST \cite{vast} & 0.2021 & 0.2358 & 0.1371 & 0.1991 & 0.2227 & 0.1311 & 0.2006 & 0.2293 & 0.1341  \\
~ & VALOR \cite{valor} & \textbf{0.8064} & \textbf{0.8375} & \textbf{0.6236} & \textbf{0.8693} & \textbf{0.8944} & \textbf{0.6848} & \textbf{0.8379} & \textbf{0.8660} & \textbf{0.6542} \\
~ & ImageBind \cite{imagebind} & 0.2803 & 0.3222 & 0.2067 & 0.0263 & 0.0223 & 0.0215 & 0.1533 & 0.1723 & 0.1141 \\
\hdashline

~ & DNN-RNT \cite{anna1} & 0.5869 & 0.7685 & 0.4147 & 0.6115 & 0.6832 & 0.4345 & 0.5992 & 0.7259 & 0.4246  \\
AVQA & DNN-SND \cite{anna2} & 0.5730 & 0.7762 & 0.4055 & 0.6162 & 0.6973 & 0.4395 & 0.5946 & 0.7368 & 0.4225 \\
~ & GeneralAVQA \cite{generalAVQA1} & 0.7143 & 0.8076 & 0.5304 & 0.6730 & 0.7201 & 0.4909 & 0.6937 & 0.7639 & 0.5107 \\
\hline

\end{tabular}
}
\vskip -0.17in
\label{tab2}
\end{table*}

\subsection{MOSs Analysis}
\label{ssec:mos}

Fig.~\ref{fig4} (a) illustrates the subjective scores of 12 SFG approaches in the two dimensions. We can observe that sonic \cite{ji2025sonic} exhibits superior overall performance, while some approaches perform poorly, likely due to resolution constraints. To further explore the effects of different categories of raw material on MOS, Fig.~\ref{fig4} (b-c) are plotted and reveal that: 1) Real human reference photographs exhibit lower visual perceptual quality than AI-generated images, yet both categories show comparable audio-visual consistency and score distributions. Further facial attribute analysis indicates severe quality degradation under two conditions: facial occlusion by accessories and profile views. 2) Musical genres show minimal score variance, whereas significant cross-lingual disparity exists. Current SFG methods demonstrate markedly lower acceptability for Chinese inputs compared to English. 3) The overall average MOS for both dimensions is only around 50, indicating that there is indeed a general quality problem in SFG content, which re-emphasizes the necessity of conducting quality assessment.

\section{benchmark experiment}
\label{sec:benchmark}

\subsection{Experimental Settings}
\label{ssec:setting}

In order to benchmark the performance of the proposed SFQA dataset, several standard quality assessment methods are selected to provide reference for future algorithm development. Specifically, the experiment included 14 evaluation methods, including:
\begin{itemize}
\setlength{\itemsep}{-3pt}
\item Traditional video quality assessment (VQA) methods: SimpleVQA \cite{sun2022simplevqa}, FAST-VQA \cite{wu2022fastvqa}, and DOVER \cite{wu2023dover}.

\item Large multi-modal model (LMM) based methods: VideoLLaMA2 \cite{damonlpsg2024videollama2} and Qwen2.5-Omni \cite{xu2025qwen25omnitechnicalreport}. We use specified prompt words to obtain dialogues that match the template for extracting results. 

\item Synchronization based methods: SyncNet \cite{Chung16syncnet} is designed to assess audio-lip consistency, yielding two key outputs: lip sync error confidence (LSE-C) and distance (LSE-D).

\item Audio-visual quality assessment (AVQA) methods: DNN-RNT \cite{anna1}, DNN-SND \cite{anna2}, and GeneralAVQA \cite{generalAVQA1}.

\item Multi-modal alignment methods: AVID-CMA \cite{avid_cma}, VAST \cite{vast}, VALOR \cite{valor}, and ImageBind \cite{imagebind}.
\end{itemize}

During the collection of raw materials, the image-audio pairs are assigned to training and validation sets, with the constraint that any individual image or audio clip does not appear in both sets simultaneously. The ratio of training set to validation set is 3.5:1, which is applicable to all methods outside Syncnet. Except for Syncnet, which directly uses the pre-trained weights, all other methods are re-trained or fine-tuned on the SFQA dataset after loading default weights, with performance evaluated using Spearman rank-order correlation (SRCC), Pearson linear correlation (PLCC), and Kendall rank-order correlation (KRCC).

\subsection{Performance \& Analysis}
\label{ssec:performance}

The experimental results are presented in Table.~\ref{tab2}, which indicating substantial scoring variance from different evaluation benchmarks. While LMM-based models and Syncnet exhibit markedly lower overall scores relative to VQA and AVQA approaches, benchmarking within multi-modal alignment methods reveals that the top-performing VALOR \cite{valor} framework surpasses other approaches by significant margins. This demonstrates the insufficient suitability of current evaluation frameworks for SFG, underscoring the critical need to develop new metrics that address challenges from SFG content.

\vspace{-0.2cm} 
\section{conclusion}
\label{sec:conclusion}

The rapid and diversified development of Singing Face Generation (SFG) has also brought about uneven quality and limited applicability scenarios. To address these challenges, this paper introduces the SFQA dataset, a quality assessment dataset specifically designed for SFG content. This dataset consists of 5,184 videos, generated by 12 representative SFG methods based on 100 reference images and 36 music clips. Through multidimensional analysis of raw material data, generation methods, and subjective experimental results, it is evident that the current SFG technology has limitations in terms of quality. Finally, this paper evaluates several existing quality assessment methods on the SFQA dataset, providing valuable benchmarks and references for the future development of new assessment methodologies.



\vfill\pagebreak


\bibliographystyle{IEEEtran}
\bibliography{IEEEabrv,strings,refs}

\begin{thebibliography}{10}
\providecommand{\url}[1]{#1}
\csname url@samestyle\endcsname
\providecommand{\newblock}{\relax}
\providecommand{\bibinfo}[2]{#2}
\providecommand{\BIBentrySTDinterwordspacing}{\spaceskip=0pt\relax}
\providecommand{\BIBentryALTinterwordstretchfactor}{4}
\providecommand{\BIBentryALTinterwordspacing}{\spaceskip=\fontdimen2\font plus
\BIBentryALTinterwordstretchfactor\fontdimen3\font minus \fontdimen4\font\relax}
\providecommand{\BIBforeignlanguage}[2]{{%
\expandafter\ifx\csname l@#1\endcsname\relax
\typeout{** WARNING: IEEEtran.bst: No hyphenation pattern has been}%
\typeout{** loaded for the language `#1'. Using the pattern for}%
\typeout{** the default language instead.}%
\else
\language=\csname l@#1\endcsname
\fi
#2}}
\providecommand{\BIBdecl}{\relax}
\BIBdecl

\bibitem{li2025samr}
Y.~Li, S.~Wu, Y.~Zhu, W.~Sun, Z.~Zhang, and S.~Song, ``Samr: Symmetric masked multimodal modeling for general multi-modal 3d motion retrieval,'' \emph{Displays}, vol.~87, p. 102987, 2025.

\bibitem{wu2023ganhead}
S.~Wu, Y.~Yan, Y.~Li, Y.~Cheng, W.~Zhu, K.~Gao, X.~Li, and G.~Zhai, ``Ganhead: Towards generative animatable neural head avatars,'' in \emph{Proceedings of the IEEE/CVF conference on computer vision and pattern recognition}, 2023, pp. 437--447.

\bibitem{wu2024mmhead}
S.~Wu, Y.~Li, Y.~Yan, H.~Duan, Z.~Liu, and G.~Zhai, ``Mmhead: Towards fine-grained multi-modal 3d facial animation,'' in \emph{Proceedings of the 32nd ACM International Conference on Multimedia}, 2024, pp. 7966--7975.

\bibitem{wu2023singinghead}
S.~Wu, Y.~Li, W.~Zhang, J.~Jia, Y.~Zhu, Y.~Yan, G.~Zhai, and X.~Yang, ``Singinghead: A large-scale 4d dataset for singing head animation,'' \emph{arXiv preprint arXiv:2312.04369}, 2023.

\bibitem{zhang2022sadtalker}
W.~Zhang, X.~Cun, X.~Wang, Y.~Zhang, X.~Shen, Y.~Guo, Y.~Shan, and F.~Wang, ``Sadtalker: Learning realistic 3d motion coefficients for stylized audio-driven single image talking face animation,'' in \emph{Proceedings of the IEEE/CVF Conference on Computer Vision and Pattern Recognition (CVPR)}, June 2023, pp. 8652--8661.

\bibitem{wei2024aniportrait}
H.~Wei, Z.~Yang, and Z.~Wang, ``Aniportrait: Audio-driven synthesis of photorealistic portrait animations,'' 2024.

\bibitem{wang2024V-Express}
\BIBentryALTinterwordspacing
C.~Wang, K.~Tian, J.~Zhang, Y.~Guan, F.~Luo, F.~Shen, Z.~Jiang, Q.~Gu, X.~Han, and W.~Yang, ``V-express: Conditional dropout for progressive training of portrait video generation,'' \emph{CoRR}, vol. abs/2406.02511, 2024. [Online]. Available: \url{https://doi.org/10.48550/arXiv.2406.02511}
\BIBentrySTDinterwordspacing

\bibitem{cui2024hallo2}
J.~Cui, H.~Li, Y.~Yao, H.~Zhu, H.~Shang, K.~Cheng, H.~Zhou, S.~Zhu, and J.~Wang, ``Hallo2: Long-duration and high-resolution audio-driven portrait image animation,'' 2024.

\bibitem{ki2024float}
T.~Ki, D.~Min, and G.~Chae, ``Float: Generative motion latent flow matching for audio-driven talking portrait,'' \emph{arXiv preprint arXiv:2412.01064}, 2024.

\bibitem{zheng2024memo}
L.~Zheng, Y.~Zhang, H.~Guo, J.~Pan, Z.~Tan, J.~Lu, C.~Tang, B.~An, and S.~Yan, ``Memo: Memory-guided diffusion for expressive talking video generation,'' \emph{arXiv preprint arXiv:2412.04448}, 2024.

\bibitem{chen2024echomimic}
Z.~Chen, J.~Cao, Z.~Chen, Y.~Li, and C.~Ma, ``Echomimic: Lifelike audio-driven portrait animations through editable landmark conditions,'' in \emph{Proceedings of the AAAI Conference on Artificial Intelligence}, vol.~39, no.~3, 2025, pp. 2403--2410.

\bibitem{ji2025sonic}
X.~Ji, X.~Hu, Z.~Xu, J.~Zhu, C.~Lin, Q.~He, J.~Zhang, D.~Luo, Y.~Chen, Q.~Lin, Q.~Lu, and C.~Wang, ``Sonic: Shifting focus to global audio perception in portrait animation,'' in \emph{Proceedings of the IEEE/CVF Conference on Computer Vision and Pattern Recognition (CVPR)}, June 2025, pp. 193--203.

\bibitem{cui2024hallo3}
J.~Cui, H.~Li, Y.~Zhan, H.~Shang, K.~Cheng, Y.~Ma, S.~Mu, H.~Zhou, J.~Wang, and S.~Zhu, ``Hallo3: Highly dynamic and realistic portrait image animation with video diffusion transformer,'' in \emph{Proceedings of the IEEE/CVF Conference on Computer Vision and Pattern Recognition (CVPR)}, June 2025, pp. 21\,086--21\,095.

\bibitem{qiu2025skyreels}
D.~Qiu, Z.~Fei, R.~Wang, J.~Bai, C.~Yu, M.~Fan, G.~Chen, and X.~Wen, ``Skyreels-a1: Expressive portrait animation in video diffusion transformers,'' \emph{arXiv preprint arXiv:2502.10841}, 2025.

\bibitem{wang2025fantasytalking}
M.~Wang, Q.~Wang, F.~Jiang, Y.~Fan, Y.~Zhang, Y.~Qi, K.~Zhao, and M.~Xu, ``Fantasytalking: Realistic talking portrait generation via coherent motion synthesis,'' \emph{arXiv preprint arXiv:2504.04842}, 2025.

\bibitem{tan2025dicetalk}
W.~Tan, C.~Lin, C.~Xu, F.~Xu, X.~Hu, X.~Ji, J.~Zhu, C.~Wang, and Y.~Fu, ``Disentangle identity, cooperate emotion: Correlation-aware emotional talking portrait generation,'' \emph{arXiv preprint arXiv:2504.18087}, 2025.

\bibitem{Meng_2025_EchoMimicV2}
R.~Meng, X.~Zhang, Y.~Li, and C.~Ma, ``Echomimicv2: Towards striking, simplified, and semi-body human animation,'' in \emph{Proceedings of the IEEE/CVF Conference on Computer Vision and Pattern Recognition (CVPR)}, June 2025, pp. 5489--5498.

\bibitem{li2025aghi}
Y.~Li, S.~Wu, W.~Sun, Z.~Zhang, Y.~Zhu, Z.~Zhang, H.~Duan, X.~Min, and G.~Zhai, ``Aghi-qa: A subjective-aligned dataset and metric for ai-generated human images,'' \emph{arXiv preprint arXiv:2504.21308}, 2025.

\bibitem{sun2022simplevqa}
W.~Sun, X.~Min, W.~Lu, and G.~Zhai, ``A deep learning based no-reference quality assessment model for ugc videos,'' in \emph{Proceedings of the 30th ACM International Conference on Multimedia}, 2022, p. 856–865.

\bibitem{wu2022fastvqa}
H.~Wu, C.~Chen, J.~Hou, L.~Liao, A.~Wang, W.~Sun, Q.~Yan, and W.~Lin, ``Fast-vqa: Efficient end-to-end video quality assessment with fragment sampling,'' in \emph{Proceedings of European Conference of Computer Vision (ECCV)}, 2022.

\bibitem{gao2025ges}
Z.~Gao, Y.~Li, S.~Wu, Y.~Cao, H.~Duan, and G.~Zhai, ``Ges-qa: A multidimensional quality assessment dataset for audio-to-3d gesture generation,'' \emph{arXiv preprint arXiv:2508.12020}, 2025.

\bibitem{zhou2026mi3s}
Y.~Zhou, Z.~Zhang, S.~Wu, J.~Jia, Y.~Jiang, W.~Sun, X.~Liu, X.~Min, and G.~Zhai, ``Mi3s: A multimodal large language model assisted quality assessment framework for ai-generated talking heads,'' \emph{Information Processing \& Management}, vol.~63, no.~1, p. 104321, 2026.

\bibitem{Chung16syncnet}
J.~S. Chung and A.~Zisserman, ``Out of time: automated lip sync in the wild,'' in \emph{Workshop on Multi-view Lip-reading, ACCV}, 2016.

\bibitem{wusijing2025fvq}
S.~Wu, Y.~Li, Z.~Xu, Y.~Gao, H.~Duan, W.~Sun, and G.~Zhai, ``Fvq: A large-scale dataset and a lmm-based method for face video quality assessment,'' \emph{arXiv preprint arXiv:2504.09255}, 2025.

\bibitem{zhou2025better}
Y.~Zhou, J.~Cao, Z.~Zhang, F.~Wen, Y.~Jiang, J.~Jia, X.~Liu, X.~Min, and G.~Zhai, ``Who is a better talker: Subjective and objective quality assessment for ai-generated talking heads,'' \emph{arXiv preprint arXiv:2507.23343}, 2025.

\bibitem{CelebAMask-HQ}
C.-H. Lee, Z.~Liu, L.~Wu, and P.~Luo, ``Maskgan: Towards diverse and interactive facial image manipulation,'' in \emph{IEEE Conference on Computer Vision and Pattern Recognition (CVPR)}, 2020.

\bibitem{sd3.5}
P.~Esser, S.~Kulal, A.~Blattmann, R.~Entezari, J.~M{\"u}ller, H.~Saini, Y.~Levi, D.~Lorenz, A.~Sauer, F.~Boesel \emph{et~al.}, ``Scaling rectified flow transformers for high-resolution image synthesis,'' in \emph{Forty-first international conference on machine learning}, 2024.

\bibitem{lei2025levo}
S.~Lei, Y.~Xu, Z.~Lin, H.~Zhang, W.~Tan, H.~Chen, J.~Yu, Y.~Zhang, C.~Yang, H.~Zhu, S.~Wang, Z.~Wu, and D.~Yu, ``Levo: High-quality song generation with multi-preference alignment,'' \emph{arXiv preprint arXiv:2506.07520}, 2025.

\bibitem{damonlpsg2024videollama2}
Z.~Cheng, S.~Leng, H.~Zhang, Y.~Xin, X.~Li, G.~Chen, Y.~Zhu, W.~Zhang, Z.~Luo, D.~Zhao \emph{et~al.}, ``Videollama 2: Advancing spatial-temporal modeling and audio understanding in video-llms,'' \emph{CoRR}, 2024.

\bibitem{xu2025qwen25omnitechnicalreport}
\BIBentryALTinterwordspacing
J.~Xu, Z.~Guo, J.~He, H.~Hu, T.~He, S.~Bai, K.~Chen, J.~Wang, Y.~Fan, K.~Dang, B.~Zhang, X.~Wang, Y.~Chu, and J.~Lin, ``Qwen2.5-omni technical report,'' 2025. [Online]. Available: \url{https://arxiv.org/abs/2503.20215}
\BIBentrySTDinterwordspacing

\bibitem{wu2023dover}
H.~Wu, E.~Zhang, L.~Liao, C.~Chen, J.~H. Hou, A.~Wang, W.~S. Sun, Q.~Yan, and W.~Lin, ``Exploring video quality assessment on user generated contents from aesthetic and technical perspectives,'' in \emph{International Conference on Computer Vision (ICCV)}, 2023.

\bibitem{avid_cma}
P.~Morgado, N.~Vasconcelos, and I.~Misra, ``Audio-visual instance discrimination with cross-modal agreement,'' 2020.

\bibitem{vast}
S.~Chen, H.~Li, Q.~Wang, Z.~Zhao, M.~Sun, X.~Zhu, and J.~Liu, ``Vast: A vision-audio-subtitle-text omni-modality foundation model and dataset,'' \emph{Advances in Neural Information Processing Systems}, vol.~36, 2024.

\bibitem{valor}
S.~Chen, X.~He, L.~Guo, X.~Zhu, W.~Wang, J.~Tang, and J.~Liu, ``Valor: Vision-audio-language omni-perception pretraining model and dataset,'' \emph{arXiv preprint arXiv:2304.08345}, 2023.

\bibitem{imagebind}
R.~Girdhar, A.~El-Nouby, Z.~Liu, M.~Singh, K.~V. Alwala, A.~Joulin, and I.~Misra, ``Imagebind: One embedding space to bind them all,'' in \emph{CVPR}, 2023.

\bibitem{anna1}
Y.~Cao, X.~Min, W.~Sun, and G.~Zhai, ``Attention-guided neural networks for full-reference and no-reference audio-visual quality assessment,'' \emph{IEEE Transactions on Image Processing}, vol.~32, pp. 1882--1896, 2023.

\bibitem{anna2}
------, ``Deep neural networks for full-reference and no-reference audio-visual quality assessment,'' in \emph{2021 IEEE International Conference on Image Processing (ICIP)}, 2021, pp. 1429--1433.

\bibitem{generalAVQA1}
{Y. Cao, X. Min, W. Sun, and G. Zhai}, ``Subjective and objective audio-visual quality assessment for user generated content,'' \emph{IEEE Transactions on Image Processing}, vol.~32, pp. 3847--3861, 2023.

\end{thebibliography}

\end{document}